\documentclass[final]{aipproc}
\layoutstyle{6x9}

\newcommand{\be}{\begin{eqnarray*}}
\newcommand{\ee}{\end{eqnarray*}}
\newcommand{\bee}{\begin{eqnarray}}
\newcommand{\eee}{\end{eqnarray}}
\newcommand{\alphas}{\alpha_s}

\newcommand{\squark}{\tilde{\rm q}}
\newcommand{\Squark}{\tilde{\rm Q}}
\newcommand{\Sup}{\tilde{\rm u}}
\newcommand{\Sdown}{\tilde{\rm d}}
\newcommand{\Scharm}{\tilde{\rm c}}
\newcommand{\Sstrange}{\tilde{\rm s}}
\newcommand{\Stop}{\tilde{\rm t}}
\newcommand{\Sbottom}{\tilde{\rm b}}

\newcommand{\pT}{{\rm p}_{\rm T}}

\begin{document}

\vspace*{-15mm}
\begin{flushright}
MPP-2009-160\\
\end{flushright}
\vspace*{0.2cm}

\title{NLO electroweak contributions to squark pair production at the LHC}

\classification{12.15.Lk}
\keywords      {MSSM, NLO Computations, Hadronic Colliders}

\author{Jan Germer}{
  address={Max-Planck-Institut f\"ur Physik, F\"ohringer Ring 6,
    D-80805 M\"unchen, Germany} 
}

\author{Wolfgang Hollik}{
  address={Max-Planck-Institut f\"ur Physik, F\"ohringer Ring 6,
    D-80805 M\"unchen, Germany}
}

\author{Edoardo Mirabella}{
  address={Max-Planck-Institut f\"ur Physik, F\"ohringer Ring 6,
    D-80805 M\"unchen, Germany}
}

\author{Maike K. Trenkel}{
  address={Max-Planck-Institut f\"ur Physik, F\"ohringer Ring 6,
    D-80805 M\"unchen, Germany}
}

\begin{abstract}
We present the tree-level and next-to-leading order (NLO) electroweak
(EW) contributions to squark--squark production at the Large Hadron
Collider (LHC) within the framework of the Minimal Supersymmetric
Standard Model (MSSM).
\end{abstract}

\maketitle


\section{Introduction}

If supersymmetry is realized at the TeV scale it will be probed at the
LHC. The direct search for colored SUSY particles, i.e. squarks and
gluinos, is of special interest owing to their large production
rate at hadron colliders.\par
As known from stop--anti-stop \cite{Hollik:2007wf},
squark--anti-squark \cite{Hollik:2008yi} and squark--gluino
\cite{Hollik:2008vm} production, NLO electroweak (EW) contributions of
$\mathcal{O}(\alphas^2\alpha)$ were found to be significant in
specific configurations, and of comparable size as the tree-level
EW contributions of $\mathcal{O}(\alphas\alpha+\alpha^2)$
\cite{Bornhauser:2007bf}. EW effects to gluino--gluino production are
only present at the loop level, and were found to be small
\cite{Mirabella:2009ap}.\par
Here we focus on the process of squark--squark production:
\bee
  {\rm PP} \rightarrow \Squark_{\alpha} \Squark_{\beta}^{\prime} + X, \quad
  (\Squark^{(\prime)} \ne \Stop, \Sbottom ),\label{process}
\eee
where $\{ \alpha, \beta \} =\{{ \rm L,R} \}$ label the chirality of the
squarks, neglecting left-right mixing. The final state squarks have to
be of the same generation as the initial state quarks. We do not
consider the production of top (bottom) squarks due to the vanishing
(small) density of the corresponding quark inside the proton. In total
one has to consider 36 distinct processes, resulting from the various
possible combinations of squarks of different flavor and chirality in
the final state.\par
In the following we will consider the tree-level and next-to-leading
order electroweak contributions to the processes~(\ref{process}). This work
is a yet missing part of our ongoing project on the computation of the
complete EW contributions up to $\mathcal{O}(\alphas^2\alpha)$ to all
squark and gluino pair production processes at the LHC.

\section{Tree-level EW contributions}
In general, squark--squark production processes can be divided into
three classes according to the flavor of the produced squarks:
\begin{enumerate}
  \item Production of two same-flavor squarks, e.g. ${\rm PP} \rightarrow
    \Sup_\alpha \Sup_\beta, \Sdown_\alpha \Sdown_\beta,$ \dots
  \item Production of two squarks of different flavor, belonging to the
    same SU(2) doublet,\\
    ${\rm PP}\rightarrow \Sup_{\alpha} \Sdown_{\beta}, \Scharm_{\alpha}
    \Sstrange_{\beta}$.
  \item Production of different squarks in different SU(2)
    doublets, e.g.\\
    ${\rm PP} \rightarrow \Sup_\alpha \Scharm_\beta, \Sup_\alpha
    \Sstrange_\beta,$ \dots
\end{enumerate}
At tree-level, squark--squark production can only be induced via
gluino, neutralino or chargino exchange in the t- or
u-channel. In the following we will refer to a diagram with gluino
(neutralino/chargino) exchange as a QCD diagram (EW diagram).
The leading order cross section of $\mathcal{O}(\alphas^2)$ as
well as a contribution of $\mathcal{O}(\alpha^2)$ is present in all
three classes through the squared QCD and EW matrix elements,
respectively. In the first two classes, one also finds a contribution of
$\mathcal{O}(\alphas\alpha)$ from the non-vanishing interference of
QCD and EW diagrams.

\section{NLO EW contributions}

The NLO EW contributions of $\mathcal{O}(\alphas^2\alpha)$
consist of virtual corrections and of real corrections from 
photon, gluon, and quark emission. Various interference terms have to
be selected carefully to get an IR finite result in the aimed order in
perturbation theory.

\subsection{Virtual corrections}
Three
 different
 types of interference terms constitute the virtual corrections at
$\mathcal{O}(\alphas^2\alpha)$:
\begin{itemize}
  \item The interference contributions of tree-level QCD diagrams with
    one-loop graphs obtained from tree-level QCD diagrams with EW insertions.
  \item The interference contributions of tree-level QCD diagrams with
    one-loop graphs obtained from tree-level EW diagrams with QCD insertions.
  \item The interference contributions of tree-level EW diagrams with
    one-loop graphs obtained from tree-level QCD diagrams with QCD insertions.
\end{itemize}
The on-shell scheme is used to renormalize the masses and the
wave functions of the quarks, squarks and gluinos
\cite{Denner:1991kt,Hollik:2003jj}. The strong
coupling constant is renormalized in the $\overline{{\rm MS}}$ scheme
with five flavor running of $\alphas$. SUSY Slavnov-Taylor identities
are restored by adding a proper counterterm for the squark--quark--gluino
coupling \cite{Beenakker:1996ch,Hollik:2001cz}.

\subsection{Real corrections}
To obtain IR and collinear finite results, the processes of real photon
and real gluon emission have to be included. At
$\mathcal{O}(\alphas^2\alpha)$, the former is given by the squared
matrix element of tree-level QCD diagrams radiating a photon, while
the latter is obtained by the interference contributions of EW and
QCD tree-level diagrams radiating a gluon. IR and collinear
singularities have been regularized using mass regularization.
The EW contribution (QCD contribution), which is given by the sum of real
photon emission (real gluon emission) and EW (QCD) loop insertions,
is IR safe. The remaining universal collinear singularities are
absorbed into the definition of the quark distribution functions (PDF).
We also consider real quark emission at
$\mathcal{O}(\alphas^2\alpha)$. At this order, only universal
collinear singularities arise that also have to be absorbed by
redefining the quark PDF. In specific scenarios, internal gluinos,
neutralinos or charginos can go on shell if they are heavier than one
of the produced squarks. If this is the case, we regularize the
singularity by including the respective particle width in the resonant
propagator.

\section{Numerical results}
Diagrams and corresponding amplitudes were generated using {\tt
  FeynArts} \cite{Hahn:2000kx,Hahn:2001rv} while the algebraic
simplifications and numerical evaluation of the scalar integrals were
performed using {\tt FormCalc} and {\tt  LoopTools} \cite{Hahn:1998yk,Hahn:2006qw}.
For illustration of the EW effects, we consider the SPS1a' point of
the MSSM, suggested by the SPA convention~\cite{Allanach:2002nj}. \par
In Table~\ref{table1} we give results for the integrated total cross
section. We refer to the three different combinations of chiralities of the
produced squarks (LL, LR, RR) and to the inclusive $\squark
\squark^{\prime}$ process, respectively.
\begin{table}
  \begin{tabular}{c|r|r|r|r|r}
    \tablehead{0}{r}{b}{}
    & \tablehead{1}{r}{b}{$\sigma^{\rm Born}_{\alphas^2}$}
    & \tablehead{1}{r}{b}{$\sigma^{\rm Tree~EW}_{\alphas\alpha+\alpha^2}$}
    & \tablehead{1}{r}{b}{$\sigma^{\rm NLO~EW}_{\alphas^2\alpha}$}
    & \tablehead{1}{r}{b}{$\delta_{\rm tree}$}
    & \tablehead{1}{r}{b}{$\delta_{\rm tree+NLO}$}\\
    \hline
    $\squark_{\rm L}\squark_{\rm L}^{\prime}$ & 1632. fb & 364. fb & -71. fb & 22.3\% & 18.0\%\\
    \hline
    $\squark_{\rm L}\squark_{\rm R}^{\prime}$ & 1682. fb & 2. fb & -69. fb & 0.1\% & -3.9\%\\
    \hline
    $\squark_{\rm R}\squark_{\rm R}^{\prime}$ & 1876. fb & 31. fb & 2. fb & 1.7\% &  1.6\%\\
    \hline
    $\squark \squark^{\prime}$ & 5189. fb & 397. fb & -141. fb & 7.7\% & 4.9\%
  \end{tabular}
  \caption{Different chiral contributions to the total hadronic cross section
    for squark--squark production at the LHC. The result is implicitly
    summed over all possible flavors $\squark, \squark^{\prime}$.
    \mbox{$\delta_{\rm tree} = \sigma^{\rm Tree~EW}_{\alphas\alpha+\alpha^2}
    / \sigma^{\rm Born}_{\alphas^2}$} and
    \mbox{$\delta_{\rm tree+NLO}= \sigma^{\rm NLO~EW}_{\alphas^2\alpha}
    / \sigma^{\rm Born}_{\alphas^2}$}.}
  \label{table1}
\end{table}
The tree-level EW contributions are positive and mainly given by the
production of two left-handed (LH) squarks with a small contribution
given by the production of two right-handed (RH) squarks. The
NLO EW corrections are negative and equally constituted by the
corrections to the production of two LH squarks and two squarks of
different chiralities. In contrast, the tree-level EW
contribution to the production of squarks of different chiralities and
the NLO EW contribution for two RH squarks is negligible. This
reflects the non-trivial behavior of the NLO EW corrections;
it is not possible to give a general correction factor to the
tree-level EW results. The impact of the EW
contributions on the inclusive total cross section reduces from
$7.7\%$ to $4.9\%$ if NLO EW corrections are included.
To illustrate the significance of EW contributions on differential
distributions we show in Figure~\ref{fig1} the differential $\pT$
distribution for $\Sup_{\rm L} \Sup_{\rm L}$ production.
\renewcommand{\tablename}{FIGURE}
\setcounter{table}{0}
\begin{table}
  \begin{tabular}{cc}
    $\stackrel{\rm \bf (a)} {\includegraphics[height=120pt,width=.45\textwidth]{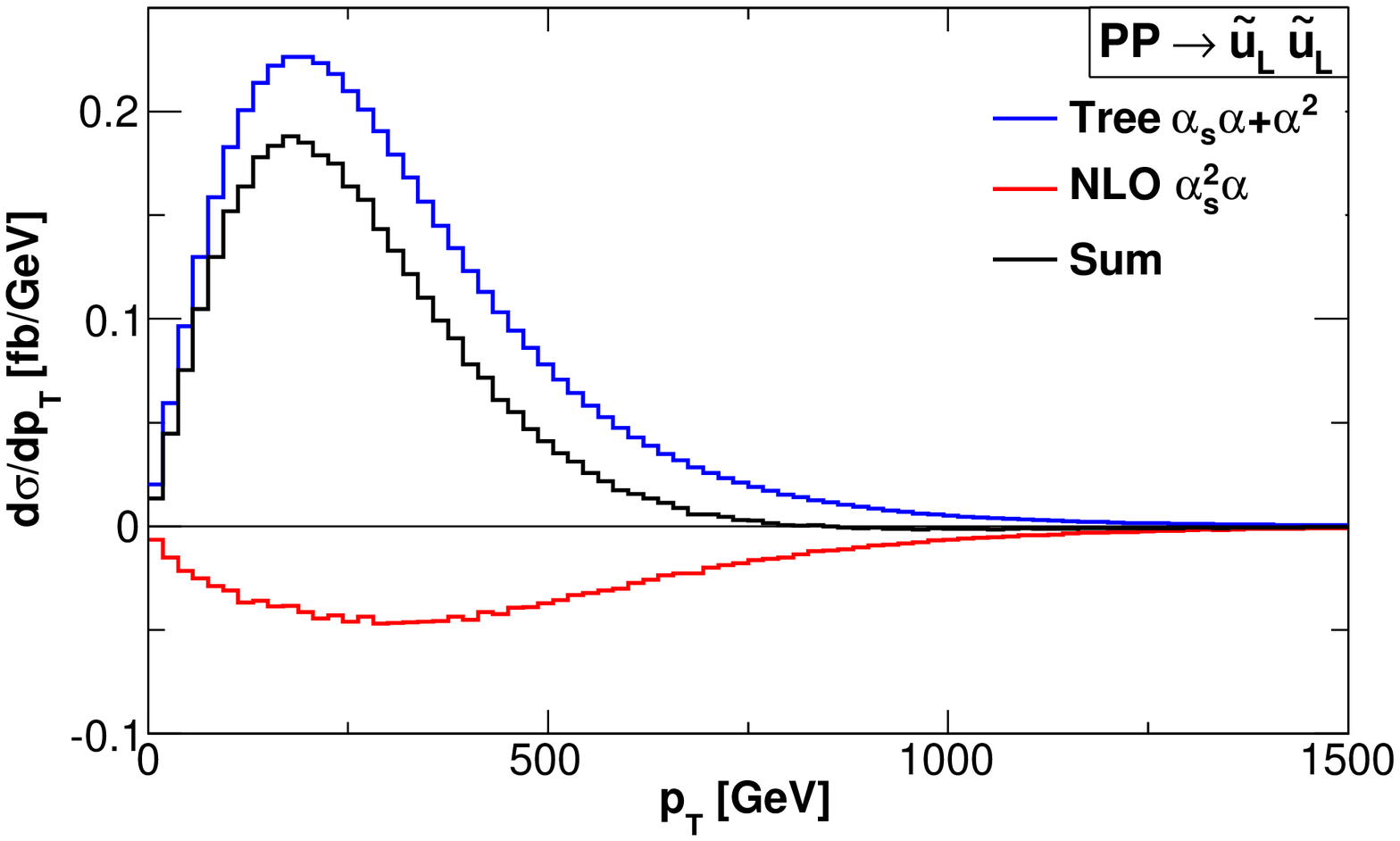}}$
    &$\stackrel{\rm \bf (b)} {\includegraphics[height=120pt,width=.45\textwidth]{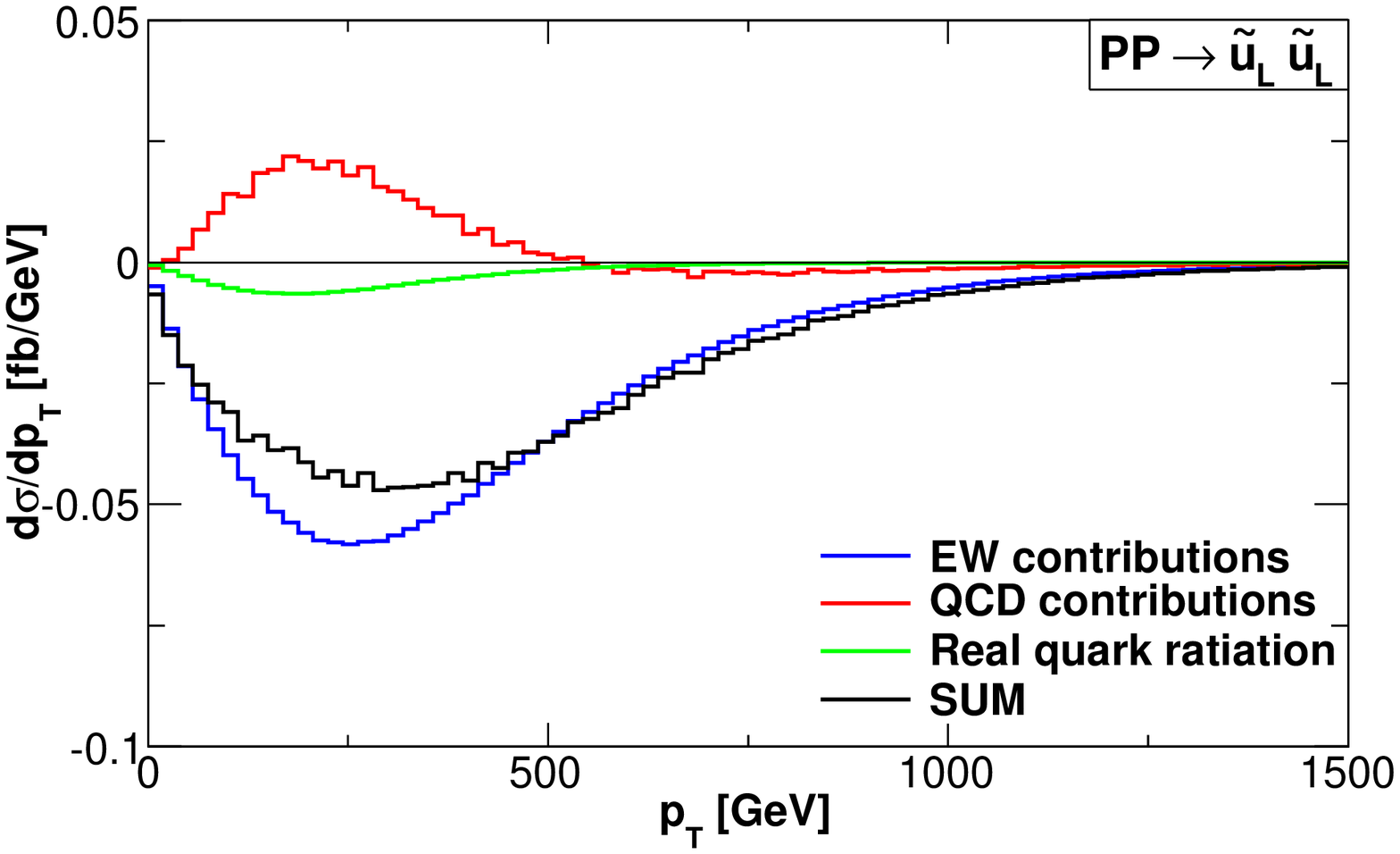}}$\\
    \parbox{.45\textwidth}{\vspace{-100pt}
      \begin{flushleft}
    \begin{description}
    \item[(a)] The tree-level EW and the NLO EW contribution, as well
      as the sum of both.
    \item[(b)] Different contributions of $\mathcal{O}(\alphas^2\alpha)$.
    \item[(c)] Relative contribution of the NLO EW corrections,
      $\delta =\mathcal{O}(\alphas^2\alpha) /
      \mathcal{O}(\alphas^2)$.
    \end{description}
    \end{flushleft}
    }
    &$\stackrel{\rm \bf (c)} {\includegraphics[height=120pt,width=.45\textwidth]{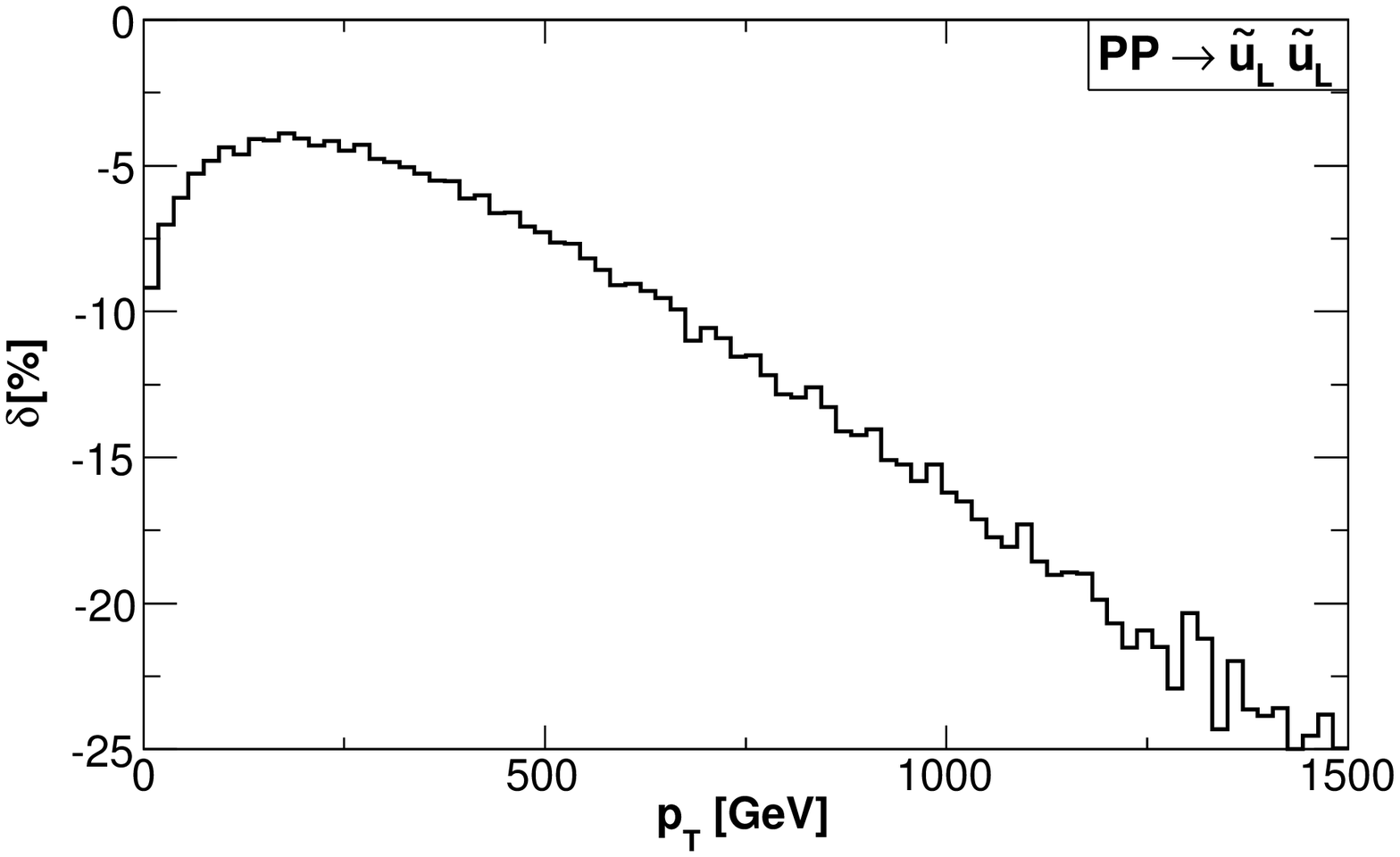}}$
    \caption{Distribution of the average $\pT$ of the squark for
      $\Sup_{\rm L} \Sup_{\rm L}$ production at the LHC.}
    \label{fig1}
  \end{tabular}
\end{table}
At low $\pT$, the EW and QCD contributions have different sign, and
therefore partially cancel in the sum. As usual, as one can see in
Figure~\ref{fig1}c, NLO EW corrections become more important in the
high $\pT$ region.


\bibliographystyle{aipproc}

\end{document}